\documentclass[a4paper]{llncs}

\usepackage{amssymb}
\usepackage{rotating, graphicx}
\usepackage{array}
\usepackage{subfigure}

\usepackage{url}
\usepackage{hyperref}
\newcommand{\keywords}[1]{\par\addvspace\baselineskip
\noindent\keywordname\enspace\ignorespaces#1}

\begin{document}

\mainmatter  

\title{Design and evaluation of a genomics variant analysis pipeline using GATK Spark tools}

%
%
\author{Nicholas Tucci\inst{1} \and Jacek Ca{\l}a\inst{2} \and Jannetta Steyn\inst{3} \and Paolo Missier\inst{4}}

\institute{Dipartimento di Ingegneria Elettronica,
	Universit\`a Roma Tre, Roma, Italy
	  \and
	  School of Computing, Newcastle University, UK}
  
%
%

\maketitle

\begin{abstract}
Scalable and efficient processing of genome sequence data, i.e. for variant discovery, is key to the mainstream adoption of High Throughput technology for disease prevention and for clinical use. Achieving scalability, however, requires a significant effort to enable the parallel execution of the analysis tools that make up the pipelines. This is facilitated by the new Spark versions of the well-known GATK toolkit, which offer a black-box approach by transparently exploiting the underlying Map Reduce architecture. In this paper we report on our experience implementing a standard variant discovery pipeline using GATK 4.0 with Docker-based deployment over a cluster. We provide a preliminary performance analysis, comparing the processing times and cost to those of the new Microsoft Genomics Services.

\end{abstract}

\section{Introduction}

The ability to efficiently analyse human genomes is a key component of the emerging vision for preventive, predictive, and personalised medicine \cite{Hood2011}. Genome analysis aims to discover genetic variants that help diagnose genetic diseases in clinical practice, or predict risk factors e.g. for certain types of cancer \cite{Tian2012}. A single exome contains about 10-15GB of data (encoded as a compressed FastQ file), while a whole genome totals up to 1TB. Depending on the specific kind of analysis, state of the art variant discovery and interpretation processes may take up to 10 hours to process a single exome. As whole-genome sequencing at population scale becomes economically affordable, personalised medicine will therefore increasingly require scalable variant analysis solutions.

With some variations, variant discovery consists of a pipeline where data flows through a number of well-understood steps, from the raw reads off the sequencing machine, to a list of functionally annotated variants that can be interpreted by a clinician. A number of algorithms, often implemented as open source and publicly available programs, are normally employed to implement each of the steps. A notable example is the GATK suite of programs from the Broad Institute \cite{VanderAuwera2013}, which forms the basis for the study presented in this paper, and is described more in detail below.

The most promising approach for improving the efficiency of the pipeline is to try and exploit the latent parallelism that may be available in some of the data as well as in the algorithms. In particular, there is increasing evidence that Hadoop-based implementations of deep genomic pipelines deployed on a cloud-based cluster can outperform equivalent pipelines that require HPC resources~\cite{Siretskiy2015}. In our own work~\cite{Cala2015} we have shown that a workflow-based implementation that runs on a public cloud infrastructure (Azure) scales better than a script-based HPC version, while providing better cost control. The prevalent approach to achieve parallelism at the level of the single program (see Sec. 1.2) involves partitioning the input to the program in such a way that multiple instances can be executed in parallel, one on each partition, with a merge step at the end. Clearly, this \textit{split-and-merge} pattern only works when the data chunks can be processed independently of one another. In such a case, existing tools can be \textit{wrapped} as part of the pattern, without modification. Recently, however, a new generation of GATK programs have been released (4.0, in beta version at the time of writing), which re-implement a number of the algorithms as Spark programs. In this approach, the task of achieving parallelism is essentially delegated to the Spark infrastructure in combination with HDFS for dataset partitioning.

In this paper we present an initial analysis of the new GATK facilities. We have implemented the reference GATK pipeline in Spark, using the new 4.0 programs when possible, and by wrapping the programs that have not been ported to Spark.
In the rest of the paper we describe this hybrid approach, report on the effort involved in deploying the pipeline both on a single-node Spark configuration and on a cluster, and present an initial performance evaluation on the Azure cloud for a variety of Spark settings, VM configurations, and cluster sizes. 

When variant discovery pipelines are used for research purposes, transparency and control over pipeline composition are important factors to consider, especially in view of the rapid advances in the tools. An example of open-source platform is the Genome Variant Investigation Platform (GenomeVIP)~\cite{Mashl01082017}, which employs GATK in addition to a number of other third party tools. On the other end of the spectrum, ``black box'' variant discovery services are now being offered, notably the new Microsoft Azure Genomics Services. Thanks to a grant from Microsoft, we were able to compare the GATK Spark approach with the new Microsoft Azure Genomics Services. We conclude that the Genomics Services are currently both faster and more cost-effective, when the Spark pipeline is deployed on the Azure cloud and the Spark processing times are translated into commercial rates. These results are preliminary, however, as GATK Spark tools are still in beta at the time of writing.

\subsection{The  Variant analysis pipeline}

We begin by describing the target pipeline in some detail. The pipeline is roughly aligned with the GATK Best Practices guidelines\footnote{\scriptsize \url{https://software.broadinstitute.org/gatk/best-practices/}} and incorporates the latest GATK 4.0 Spark tools. Broadly speaking, it consists of three main phases, as indicated in Fig.~\ref{fig:NGS_Pipeline_GATK}, namely \textit{Pre-processing}, \textit{Variant Discovery}, and \textit{Call Set Refinement}. The pre-processing phase takes the input raw exome dataset, in the FASTQ format, it aligns its content (unmapped reads of gene base pairs) against a reference genome like h19 or h38, using the well-known BWA aligner~\cite{Li2010a}, and it marks any duplicates, i.e., by flagging up multiple paired reads that are mapped to the same start and end positions. These reads often originate erroneously from DNA preparation methods. They will cause biases that skew variant calling and hence should be removed, in order to avoid them in downstream analysis. The BQSR (Base Quality Score Recalibration) step then assigns confidence values to each of the aligned reads, taking into account possible sequencing errors.\footnote{\scriptsize \url{https://software.broadinstitute.org/gatk/documentation/article.php?id=11081}} Finally, Variant Calling, performed using the GATK Haplotype Caller, identifies both single-nucleotide polymorphisms (SNPs) as well as insertion/deletion mutations (Indels).

Multiple variant files (gVCF), one for each sample, are then bundled together for the next phase, \textit{Variant Discovery}. The specific steps include producing raw SNP and Indel VCF files, building recalibration models for those SNPs and Indels\footnote{\scriptsize \url{https://software.broadinstitute.org/gatk/documentation/article.php?id=2805}} and refining the genotypes, that is, filtering out genotypes with low estimated accuracy. The final phase, \textit{Variant Annotation}, is not part of the Best Practices and thus may be implemented using a variety of third party tools. We used Annovar, a well-known tool for functionally annotating genetic variants detected from diverse genomes~\cite{doi:10.1093/nar/gkq603}. As mentioned later, pre-processing time dominates the entire processing time and thus our performance analysis ignores phases two and three. However, in the following we highlight some of the implementation challenges for these steps.

\subsection{Related work}  \label{sec:related}

SparkSeq~\cite{Wiewiorka2014}  is a general-purpose library for genomic cloud computing built on top of Spark. Its strengths are its generality and extensibility, as it can be used to build customised analysis pipelines (in Scala). It appears that the library is built from the ground up, i.e., without leveraging existing implementations such as GATK.

In contrast, a general big data platform for genome data analysis, called Gesall, that uses a wrapper approach to reuse existing tools without change is presented in \cite{Mushtaq2015}. Gesall leverages the potential parallelism that is available from some of the existing tools, for instance BWA, by partitioning its input (SAM and BAM files) and then managing the parallel execution of multiple BWA instances. Making this work, however, requires a heavy stack of new MapReduce-based software to be injected between the data layer (HDFS) and the native tools.

A similar approach, namely to segment input data sets and then feed them to multiple instances of the tools, is presented in  \cite{Mushtaq2015}. The distinctive element of the resulting framework is to perform load balancing by dividing chromosomal regions according to the number of reads mapped to each chromosome, as opposed to natural chromosome boundaries. 
This equalizes the size of each data chunk and, in addition to in-memory data management, achieves substantial speedup over a functionally equivalent but naively implemented Hadoop MapReduce based solution. The advantages of in-memory processing for efficient genome analysis have also been demonstrated recently in other ad hoc frameworks~\cite{Li:2018:HGA:3200691.3178511}. Yet another parallel version of a genomics pipeline that operates by partitioning the input data files is described in~\cite{Roy2017}. In this instance, however, some of the tools have been re-implemented (as opposed to simply wrapped) to explicitly leverage the embarrassingly parallel steps of the pipeline.

In contrast to these efforts, in our experiments we aim to show the potential of the tool re-implementation approach offered by the GATK 4.x tool suite, which are being incrementally ported to the Spark architecture.

\section{Spark hybrid pipeline implementation} \label{sec:implementation}

As mentioned, the main motivation for undertaking this work has been to experiment with a Spark implementation of the GATK Best Practices pipeline, based on the recently release of GATK 4.0. Not only are these tools natively built for Spark, but also, compared to the previous version (GATK 3.8), they are also better integrated with each other, for instance to avoid writing intermediate files to disk and increasing efficiency.

\begin{figure}
	\begin{center}
		\centering
		\includegraphics[width=.7\textwidth]{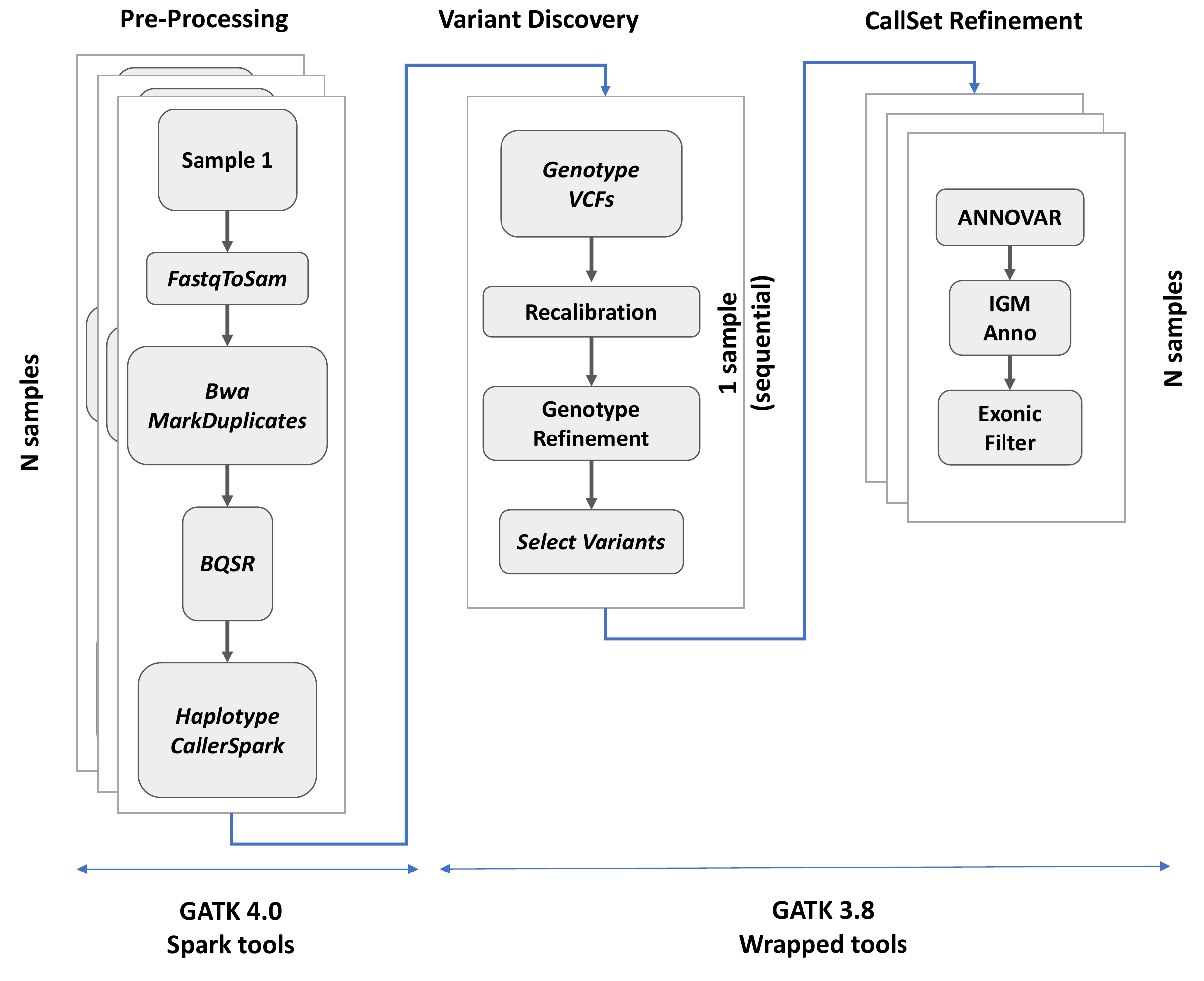}
		\caption{Multi-sample Variant processing pipeline~\label{fig:NGS_Pipeline_GATK}}
	\end{center}
\end{figure}

\sloppy At the time of writing, however, these new versions of the tools are limited to the pre-processing phase: \texttt{BwaAndMarkDuplicatesPipelineSpark}, \texttt{BQSRPipelineSpark} and \texttt{HaplotypeCallerSpark} (Fig. \ref{fig:NGS_Pipeline_GATK}). Thus, the implementation necessarily required a hybrid approach, whereby pre-processing used the new Spark tools, while for the rest of the pipeline we used a wrapper method. For this, Spark offers a transformation called \texttt{Pipe}, which ``pipes each partition of the RDD through a shell command, e.g. a Perl or bash script. RDD elements are written to the process's stdin and lines output to its stdout are returned as an RDD of strings''. Thus, \texttt{Pipe} allows Bash scripts to execute from within Spark, but not efficiently, as pipelining across the steps requires the content of intermediate RDDs to be written out to files and then be read back in. Looking at Fig.~\ref{fig:NGS_Pipeline_GATK}, it should be clear that the variant discovery phase is a potential bottleneck, as it must process the entire batch of samples, with no parallelism available. However, as it turns out its processing time is negligible compared to that of pre-processing.

\subsection{Single node deployment}

\sloppy  The hybrid native Spark/wrapper approach works well for a single-node deployment, as the entire pipeline can be launched using a single bash script that encapsulates the communication with the Spark driver. For a batch of $N$ samples, the \texttt{spark-submit} command spawns one iteration per sample for the pre-processing (\texttt{BwaAndMarkDuplicatesPipelineSpark}, \texttt{BQSRPipelineSpark}, and \texttt{HaplotypeCallerSpark}), followed by a single \texttt{VariantDiscovery} and \texttt{CallsetRefinement} call for the entire batch. The results produced by the execution have been validated against those obtained from our more established, workflow-based pipeline as described in~\cite{Cala2015}.
 
\subsection{Cluster deployment}  \label{sec:cluster}

In theory, Spark is designed to facilitate the seamless scaling out of applications over a cluster, with virtually no change to the code. The pre-processing phase of our pipeline would benefit the most from distribution, as it consists of native Spark applications as explained earlier. In reality, the deployment of a complex multi-tool pipeline like the one described requires substantial additional effort, mainly due to the requirement for Spark tools to read input and reference datasets from a HDFS data layer.

Commercial solutions such as Microsoft Azure \textit{HDInsight} provide a preconfigured environment ready to execute Spark in cluster mode. This comes at a substantial cost, however (about twice the cost of an un-configured set of VMs). We therefore undertook the challenge of a manual Spark cluster configuration.
In this section we report on our experience realising a distributed version of the pipeline using a virtualisation approach, based on \textit{Docker Swarm} technology.\footnote{\scriptsize \url{https://docs.docker.com/engine/swarm/}} Our conclusion is that while Swarm greatly simplifies deployment, manual effort is still required especially to satisfy the data access requirements of the various components, and limitations are incurred for the fragments of the pipeline that are implemented using the wrapper method as explained earlier.
Also, a distributed deployment is not always beneficial due to the additional communication overhead associated with a distributed execution, as we show in Sec.~\ref{sec:experiments}.

Swarm extends Docker by providing seamless and automated distribution of Docker containers over a cluster of VMs. A \textit{swarm} is a group of machines \textit{nodes}, that run Docker containers and are joined into a cluster. The usual Docker commands are executed on a cluster by a Swarm Manager.

Swarm managers may employ several strategies to run containers, such as ``emptiest node'', which fills the least utilized machines with containers, or ``global'', which ensures that each machine gets exactly one instance of the specified container. Swarm managers are the only machines in a swarm that can execute user commands, or authorize other machines to join the swarm as workers. Workers only provide capacity and do not have the authority to tell any other machine what it can or cannot do. In this context, a \textit{service} is an image for an application that resides in a container and that is deployed over a swarm.

\begin{figure}
	\centering
	\includegraphics[width=0.7\linewidth]{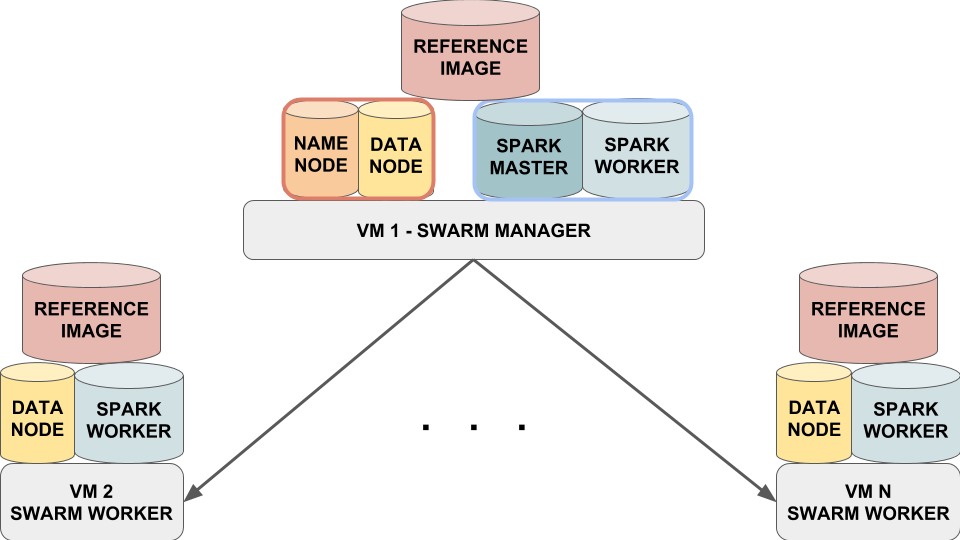}
	\caption{Virtualised Spark and HDFS cluster deployment using Docker Swarm}
	\label{fig:stacksparkhdfs}
\end{figure}

We have used Docker Swarm to deploy both Spark and HDFS over a cluster of nodes, using Docker Hub and Docker Images provided by Big Data Europe\footnote{\scriptsize \url{https://www.big-data-europe.eu/}}, as follows. The first step is to create a Swarm, which in our test cluster consists of three nodes: a Swarm Manager and two Swarm Workers as shown in Fig.~\ref{fig:stacksparkhdfs}. As both Spark and HDFS adopt Master-Slave architecture, the masters (Spark Master and HDFS Namenode) are deployed on the Swarm Manager. The Slaves (Spark workers and HDFS Data nodes) are deployed globally, that is, one replica is allocated to each node in the Swarm, including the Swarm Manager node. The Docker containers that host these images are connected through a dedicated overlay network.

Shared data, including all input samples, reference databases, GATK libraries, etc., resides on HDFS and is therefore naturally distributed and replicated over the Data nodes across the cluster. For the most part, this achieves location transparency as tools need only access the data through Spark HDFS drivers (readers and writers). There are two exceptions, however. Firstly, nonSpark tools expect data to be accessible on a local file system. This is achieved by mounting HDFS Data nodes as virtual Docker volumes so they are accessible from within a Docker container. Secondly, the reference genome had to be replicated to each local Worker file system (see \textit{reference image} in Fig.~\ref{fig:stacksparkhdfs}). This is achieved by encapsulating the dataset itself as a Docker Image container, which is then automatically deployed by Swarm using the ``global'' Swarm mode, as indicated above. One advantage of this encapsulation approach is that it makes it easy to upgrade the reference genome, eg from h19 to h38.p1, the most recent.

\subsection{Cluster mode pipeline execution}  \label{sec:cluster-exec}

A key observation, already made earlier, is that none of the non-Spark programs that make up the pipeline can be distributed. This is the case for the initial step, \texttt{FastqToSam}, as well as for all the steps after pre-processing, which are necessarily executed on the Spark Master container. As the processing time is linear in the number of samples, this justifies allocating a larger VM to the Spark Master.

\begin{figure}
	\centering
	\includegraphics[width=0.7\linewidth]{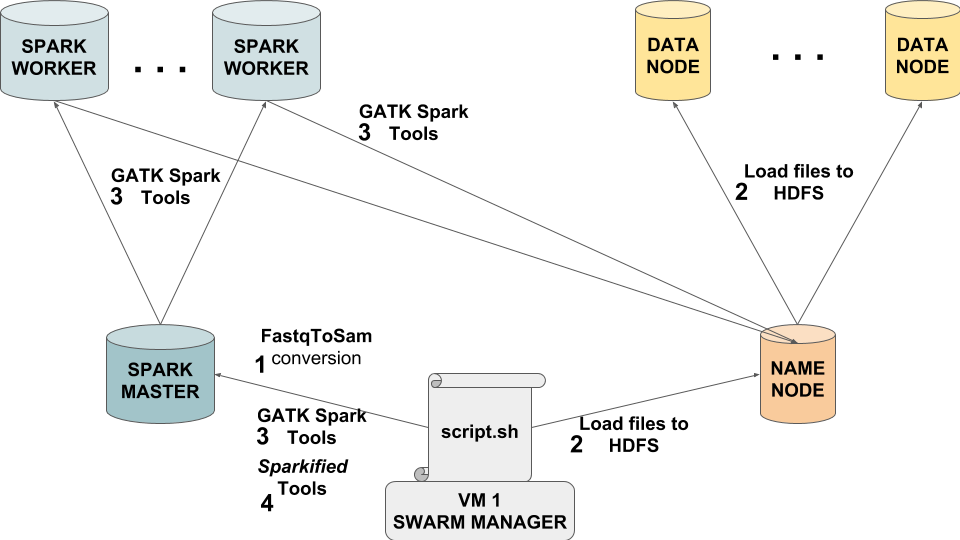}
	\caption{Pipeline execution flow in cluster mode}
	\label{fig:distributedpipeline}
\end{figure}

With this in mind, execution on a cluster consists of four main steps, controlled by a master bash script. These are summarised in Fig.~\ref{fig:distributedpipeline}. The first step, \texttt{FastqToSam}, is non-Spark and produces local uBAM files, which then needs to be distributed across the HDFS nodes (step 2) to be made available to the Spark pre-processing tools (step 3). As explained, these tools communicate through HDFS files and at the time of writing are not easy to integrate more deeply, i.e., by sharing intermediate datasets using Spark process memory. Finally, step 4 consists of the execution of non-Spark tools, again on the Spark Master. This requires that outputs that reside on HDFS be moved back to local file system.

In summary, the deployment may benefit from a partial porting of GATK tools to Spark, however non-GATK tools that escape this porting effort represent bottlenecks. Firstly, because they run in centralised mode, and secondly because of the different file infrastructure they require. Also, Spark tools appear to be designed in isolation, without attempting to eliminate intermediate data passing through HDFS reads and writes.

\section{Experimental evaluation}  \label{sec:experiments}

In this section we report on preliminary results on the performance of the pipeline. For these experiments we used 6 exomes, from anonymised patients obtained from the Institute of Genetic Medicine at Newcastle University. These samples come with naturally slightly different sizes. Our samples sizes are in the range 10.8GB-15.9GB, with average 13.5GB (compressed). Using these samples, we analysed the runtime of the pipeline implementation described in Sec.~\ref{sec:implementation}, comparing the deployment modes described in the previous section, namely a
single-node Spark model, known as ``pseudo-cluster'' mode, with a cluster mode configuration with up to four nodes. In both cases, all nodes are identical virtual machines on the Azure cloud with 8 cores, 55GB RAM. Our experiments aim to compare the effect of various Spark settings for each of these configurations.

We focused exclusively on the pre-processing phase, where the bulk of the processing occurs. Specifically, BWA alignment and duplicate marking (denoted BWA/MD in the following) accounts for 38\% of the processing time, Base Quality Score Recalibration Processing (BQSRP) for 11\%, and variant calling using the Haplotype Caller (HC) 39\%. The rest of the pipeline, which only accounts for 12\% of the processing, was not considered further in these experiments.


Four settings were used to tune the Spark configuration, indicated in the charts as X/Y/W/Z, where X is the driver process memory, Y the number of executors, W the number of cores allocated to each executor, and Z the memory allocated to each executor. 

\begin{figure}
\centerline{
	\subfigure[Configuration 20/2/4/16I]{
			\includegraphics[width=.5 \linewidth]{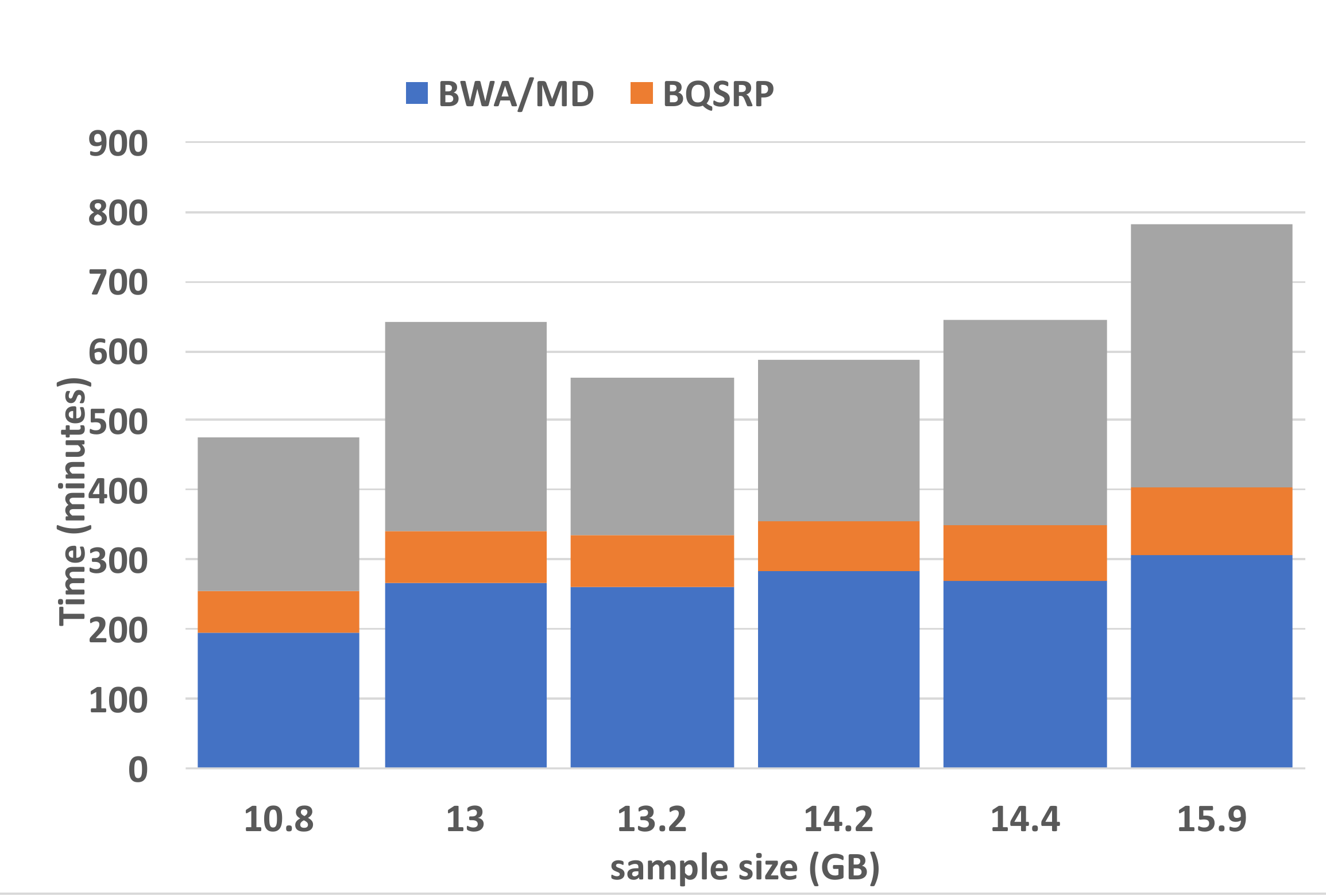}
	\label{fig:pre-proc-1}}
\quad
    \subfigure[Configuration 20/4/2/8]{
	  \includegraphics[width=.5 \linewidth]{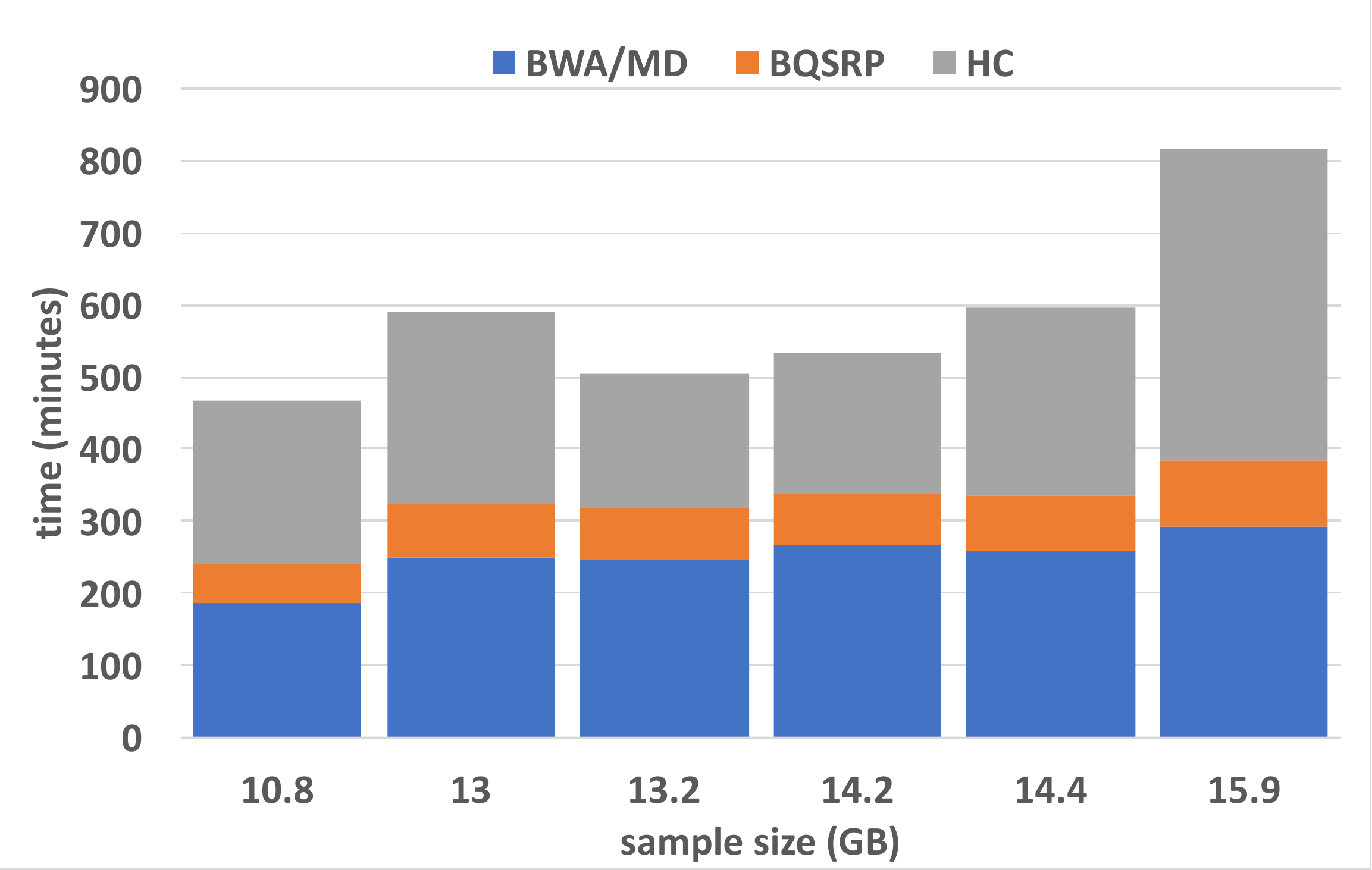}
	\label{fig:pre-proc2}}
}
\caption{Pre-processing steps for single node deployment configurations}
\label{fig:pre-proc}
\end{figure}

%

\begin{figure}
	\centerline{
		\subfigure[Average time/ GB for two configurations]{
	\includegraphics[width=0.25\linewidth]{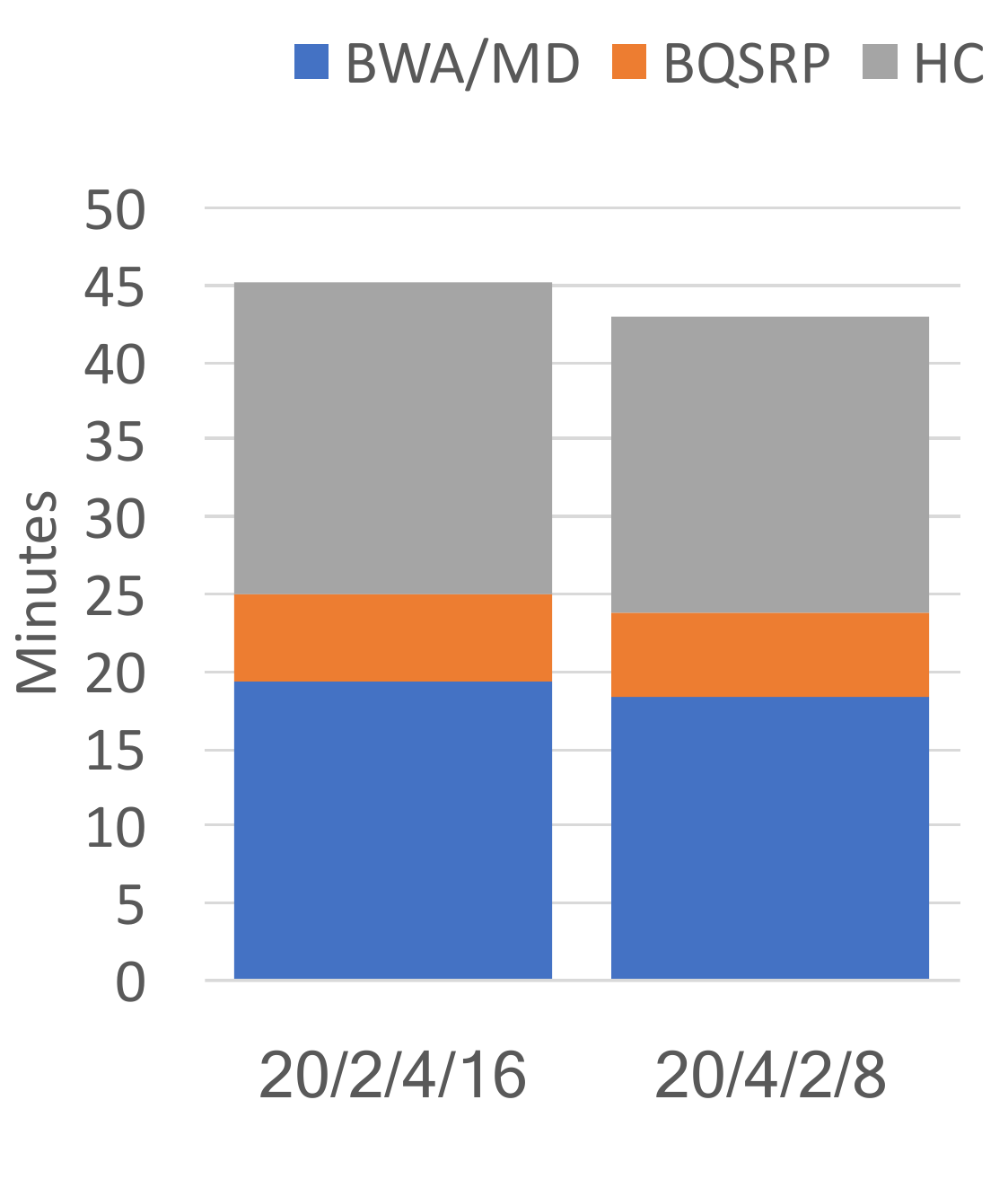}
	\label{fig:avg-time-2-config}}
		\quad
		\subfigure[Pre-processing time/GB (all three steps) across four configurations for the same sample]{
	\includegraphics[width=0.5\linewidth]{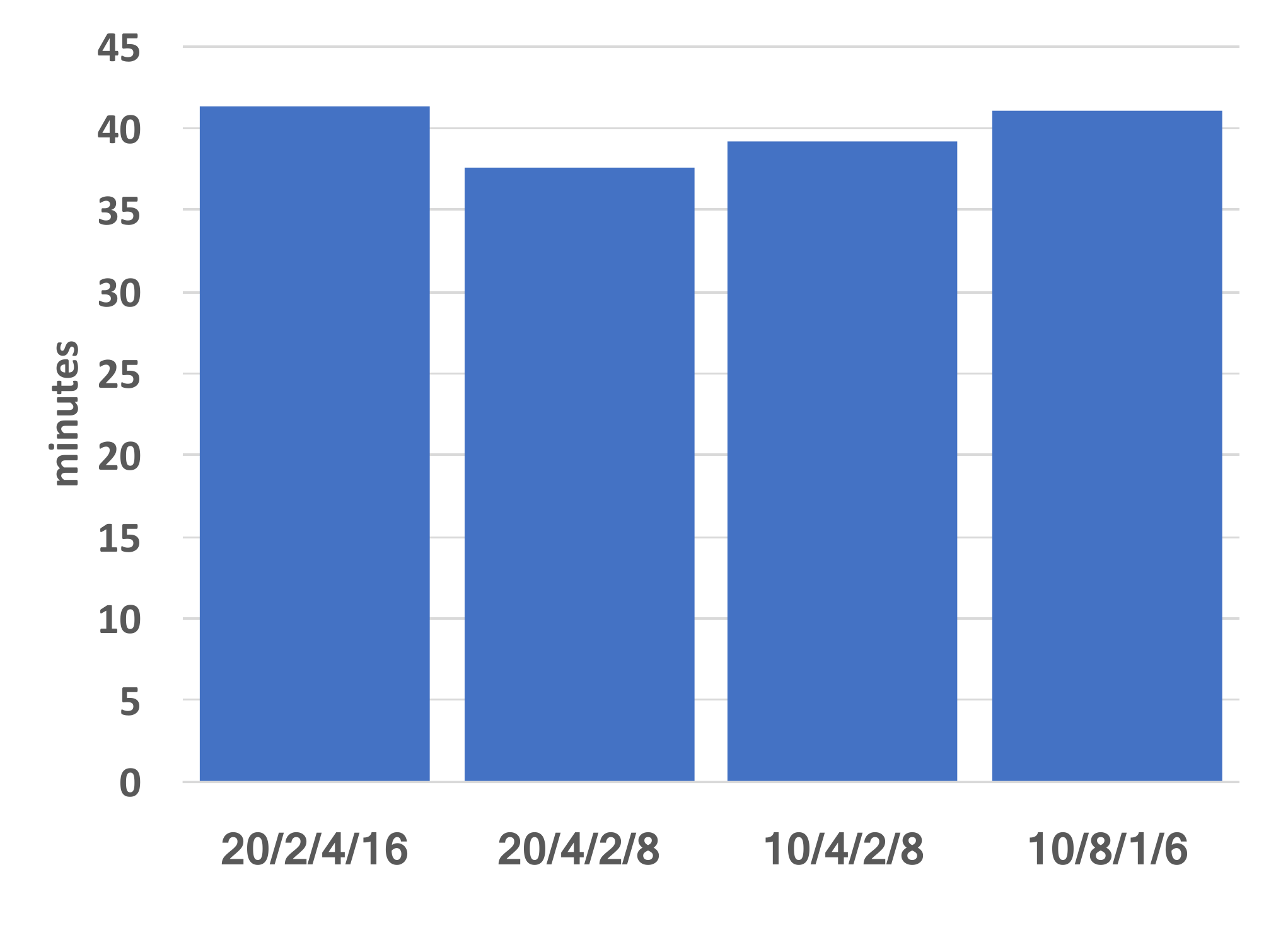}
	\label{fig:pre-proc-across-config}}
	}
	\caption{Normalised pre-processing processing time/GB }
	\label{fig:pre-across-config}
\end{figure}

Charts \ref{fig:pre-proc-1} and \ref{fig:pre-proc2} show the processing for two configurations: 20/2/4/16 and 20/4/2/8 respectively, for each of the six samples (ordered by size) and with a breakdown for each pre-processing tool. Both charts show a slight increase in processing time as the sample size increases (with an unexplained anomaly on the 13GB sample in both cases). These times are not significantly affected by the differences in configuration. Indeed, if we normalise the processing time by the input size, we observe very similar figures across the two configurations and for each tool, as shown in Fig.~\ref{fig:avg-time-2-config}. Specifically, for the two configurations BWA/MD, BQSRP, and HC report an average of 19.3 vs 18.4 minutes/GB, 5.6 vs 5.3 minutes, and 20.2 vs 19.14 minutes, respectively.

For a deeper analysis on the effect of Spark settings, we then ran the pipeline on one single representative sample (PFC 0028, 14.2GB) on two additional settings, 10/4/2/8, and 10/8/1/6. Fig. 5(b) shows the results, with processing times normalised by sample size for ease of comparison with the previous chart. Again, there is no indication that these four settings are critical in affecting the processing times.

\begin{figure}
	\centerline{
		\subfigure[single node (55GB RAM/core)]{
	\includegraphics[width=0.5\linewidth]{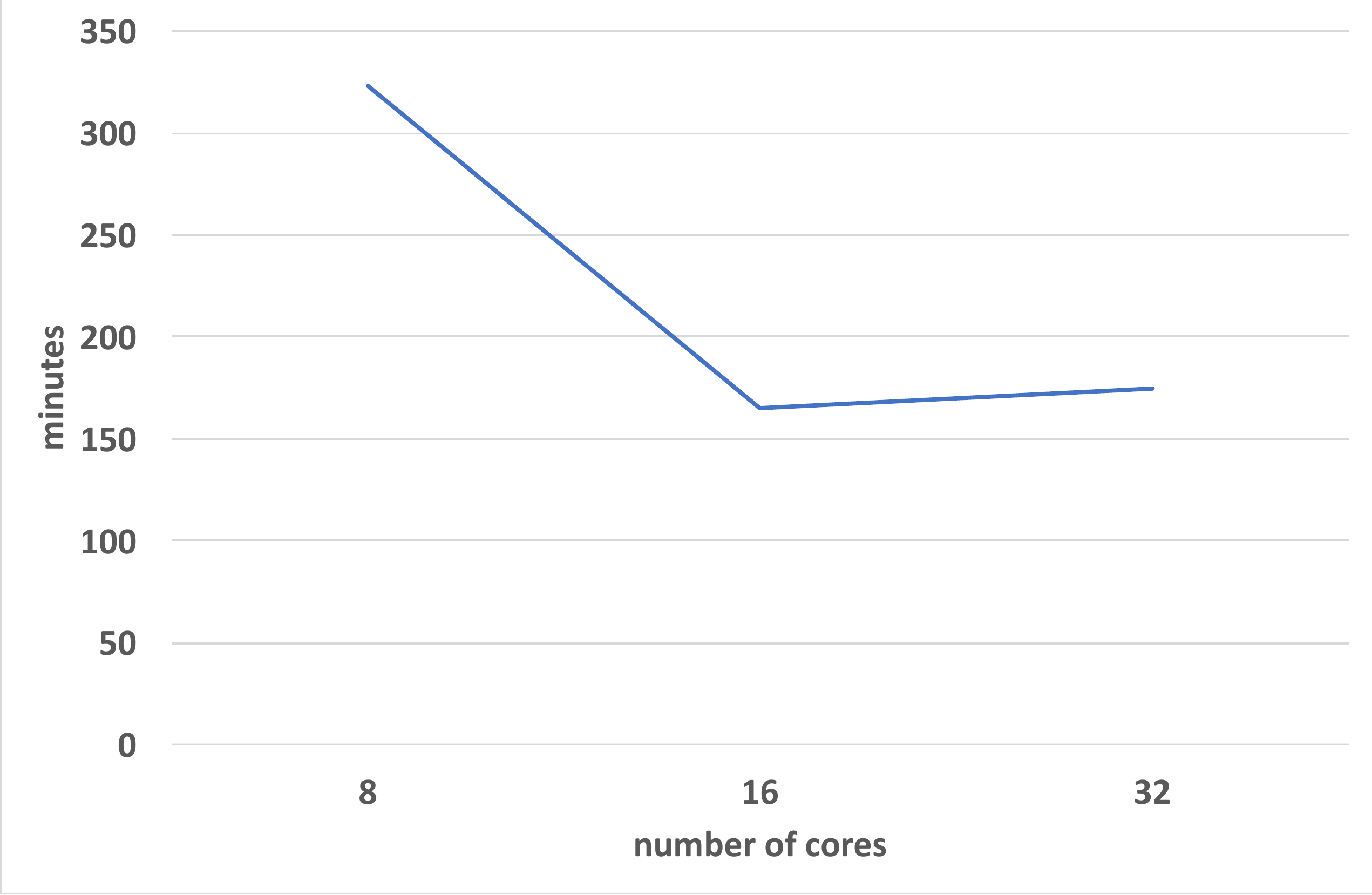}
	\label{fig:scale-up}}
		\quad
		\subfigure[8 cores/55GB RAM cluster mode]{
	\includegraphics[width=0.5\linewidth]{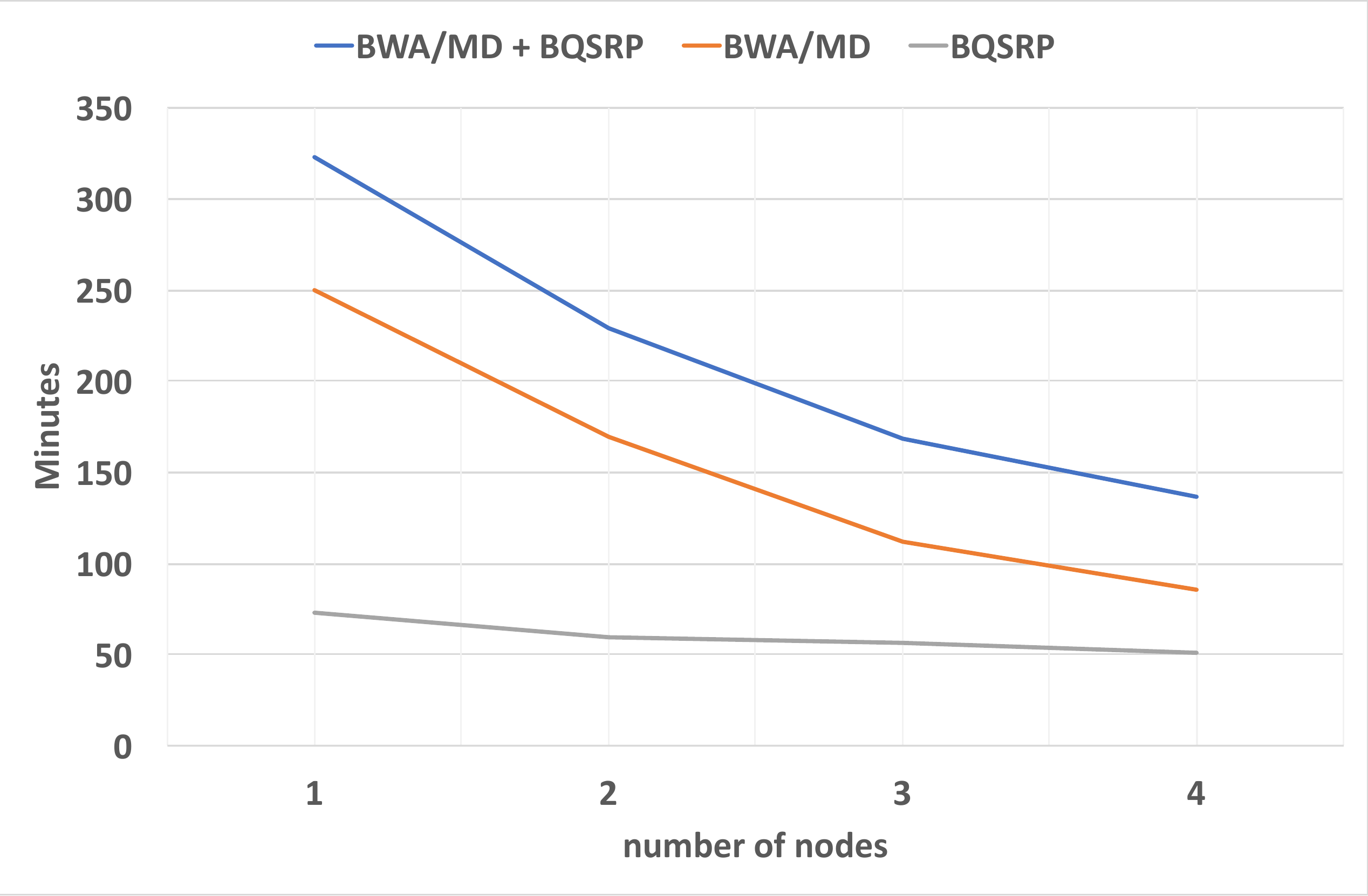}
	\label{fig:scale-out}}
	}
	\caption{BWA/MD + BQSRP speedup}
	\label{fig:speedup}
\end{figure}

More significant is the difference in processing time achieved by adding resources to the VMs. Fig.~\ref{fig:scale-up} shows a nearly ideal speedup as we double the number of cores (with a constant 55GB RAM per 8 cores, i.e. 110GB for 16 cores, etc.) It seems however that the Spark tools will not benefit from a larger VM beyond 16 cores. Note that the chart in Fig.~\ref{fig:scale-up} does not include the processing time for HC, as this took an unusually long time to run on a 16 cores configuration. This was due to an issue with a low-level library on the HC implementation, which was not resolved at the time of writing.

%
%
As expected, running Spark in cluster mode shows a speedup as we increase the number of nodes, as shown in Fig.~\ref{fig:scale-out}. However, we also note that scaling out, that is, by adding nodes, incurs an overhead that makes it less efficient than scaling up (i.e., adding cores to a single node configuration). For instance, 2 nodes with 8 cores each process at 229 minutes, while a single node with 16 cores takes 165 minutes. This overhead effect is noticeable when using 32 cores, which as we noted earlier does not improve processing time on a single host (175 minutes, Fig.~\ref{fig:scale-up}), while a 4x8 nodes cluster takes 168 minutes, a further improvement over the 2x8 configuration.

\subsection{Comparing with Microsoft Genomics Services}

Thanks to a grant from Microsoft Azure Research, we were able to process our patient samples using the new Microsoft Genomics Services. These services execute precisely the pre-processing steps of the pipeline, making it easier to compare with our results. The processing time for our reference PFC 0028 sample is an impressive 77 minutes (compare with the best time of 446 minutes on a single node, obtained from the figures in Fig.~\ref{fig:scale-up}) to which the average HC processing time has been added). However, at the time of writing these services were only offered as a \textit{black box} that runs on a single, high-end virtual machine of undisclosed specifications. In terms of pricing, the current charges for using Genomics Services are \pounds0.217 / GB, which translates to about \pounds18.61 for processing our six samples. For comparison, the cost of processing the same samples using our pipeline with a 8 cores, 55GB configuration is estimated at \pounds28.

\section{Conclusions}

We have presented an experimental evaluation of the design effort involved in implementing a genomics variant discovery pipeline using the recently released GATK Spark tools from the Broad Institute, and a performance analysis based on a single node and small cluster configuration. Our analysis is preliminary, as the GATK 4.x tools are still very recent, non-GATK tools or those that have not yet been ported represent bottlenecks. Firstly, because they run in centralised mode, and secondly because of the different file infrastructure they require. Also, Spark tools appear to be designed in isolation, without attempting to eliminate intermediate data passing through HDFS reads and writes.

Compared with the processing times reported for the Microsoft Azure Genomics Services, it appears that using Spark with the current beta version of GATK tools is currently not economically competitive and thus is not recommended for operational use in clinical settings. This may change, however, as the GATK Spark tools mature. On the plus side, our implementation offers complete control over the evolution of the pipeline over time, a key requirement especially in a genetic research setting.

\section*{Acknowledgments}
The authors are grateful to Microsoft for the Azure for Research grant that made it possible to experiment with Azure Genomics Services.


\end{document}